# From Čerenkov Radiation to Generalized Super-Čerenkov Exotic Decays

D. B. Ion[1,2] and M. L. D. Ion[3]

[1] National Institute for Physics and Nuclear Engineering Horia Hulubei, IFIN-HH, Bucharest, P.O.Box MG-6, Magurele Romania

[2] TH-Division, CERN, CH-1211 Geneva 23, Switzerland

[3] Faculty of Physics, Bucharest University, Bucharest, Romania

## Abstract

Generalized Super-Čerenkov Radiations (SČR), as well as their SČR-signatures are investigated. Two general SČR-coherence conditions are found as two natural extremes of the same spontaneous particles decays in (dielectric, nuclear or hadronic) media. The main results on the experimental test of the super-coherence conditions, obtained by using the experimental data from BNL, are presented. The interpretation of the observed anomalous Čerenkov rings as experimental evidence for the HE-component of the SČR is discussed.

## 1. Introduction

The classical theory of the radiation emitted by charged particles moving with superluminal velocities were traced back to Heaviside [1]. In fact, Heaviside considered the Čerenkov radiation [2] in a nondispersive medium. He considered this topic many times over the next 20 years, deriving most of the formalism of what is now called Čerenkov radiation and which is applied in the particle detectors technics (e.g., RICH-detectors). So, doing justice (see the papers of Kaiser and Jelley in Nature) to Heaviside [1] De Coudres [3] and Somerfeld [4], we must recall that the classical theory of the CR phenomenon in a dispersive medium was first formulated by Frank and Tamm in 1937 [5]. This theory explained all the main features of the radiation observed experimentally by Čerenkov [2] (see Fig. 1) . In fact, from experimental point of view, the electromagnetic Čerenkov radiation was first observed in the early 1900's by the experiments developed by Marie and Pierre Curie when studying radioactivity emission. In essence they observed that phenomenon consists from *the very faint emission of a bluish-white light from transparent substances in the neihbourhood of strong radioactive source.* But the first deliberate attempt to understand the origin of this light was made by Mallet [6] in 1926-1929. He observed that this *light emitted by a variety of transparent bodies placed close to a radioactive source always had the same bluish-white quality, and that the spectrum was continuous, not possessing the line or band structure characteristic of fluorescence.* Mallet found such an emission, he could not offered an explanation for the nature of this phenomenon, but was the first to find that spectrum was continuous and extended to 3700 Å. He was the first to appreciate the universality of this effect but no attempted to study the polarization of this new kind of radiation.

Only the exhaustive experimental work, carried out between years 1934-1937 by P. A. Čerenkov [2], characterized completely this kind of radiation. These experimental data are fully consistent with the classical electromagnetic theory developed by Heaviside in 1888 [1] and Frank and Tamm [5] (see Fig.1).

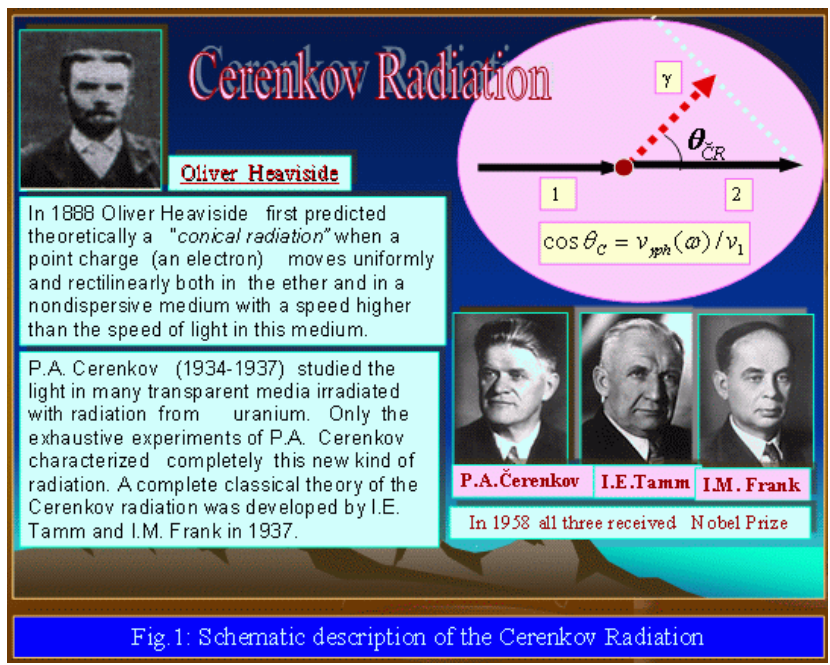

Fig.1: Schematic description of the Cerenkov Radiation

In essence, it was revealed by the Heaviside, Čerenkov, Tamm and Frank that a charged particle moving in a transparent medium with an refractive index, $n_\gamma$, and having a speed $v_1$ greater than phase velocity of light $v_{\gamma ph} = n_\gamma^{-1}$ will emit electromagnetic radiation, called Čerenkov radiation (ČR), at an polar emission angle $\theta_C$ relative to the direction of motion given by the relation (we adopted the system of units $\hbar = c = 1$) (see Fig. 2).

$$\cos\theta_C = {v_{\gamma ph}(\omega)} \Big/ {v_x} \leq 1 \qquad (1)$$

(we adopted the system of units $\hbar = c = 1$). The geometry of Čerenkov cone is presented in Fig.2.
The remarkable properties of the Čerenkov radiation find wide applications in practice especially in high energy physics where it is extensively used in experiments for counting and identifying relativistic particles [via Ring Imaging Čerenkov (RICH)-Detectors,e.g. see: T. Ypsilantis and J. Seguinot, *Nucl. Instrum. Methods A* 433, 1 (1999)] in the fields of elementary particles, nuclear physics and astrophysics. For some predictions on the *gravitational Čerenkov-like effects* see: M. Pardy, Phys. Lett. **B 336**, 362 (1994); Int. J. Theor. Phys. **34**, 951 (1995), **41**, 887 (2002); A. B. Balakin et al., Class. Quant. Grav. **18**, 2217 (2001).
A quantum theoretical approach of the Čerenkov effect by Ginsburg [7] resulted in only minor modification to the classical theory [see also the books [8-9]].

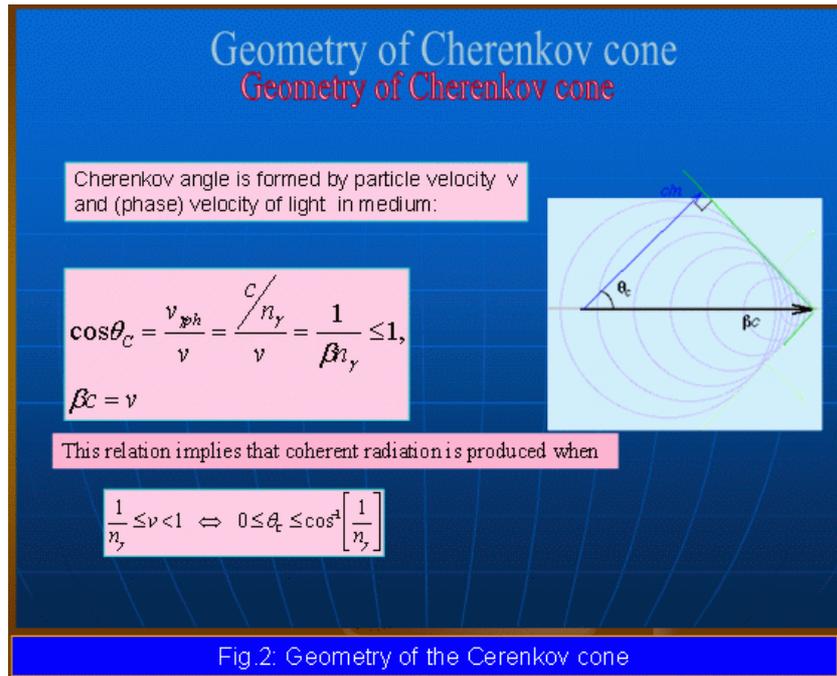

Fig.2: Geometry of the Cerenkov cone

Some interesting discussions about the predictions and experimental discovery of the Čerenkov radiation can be found in the papers of Kaiser [10], Jelley [11], Tyapkin [12] and Govorkov [13].
In Fig. 3. we can see the Čerenkov blue light emitted by the electrons which are traveling faster than speed of light in water surrounding a nuclear reactor core.

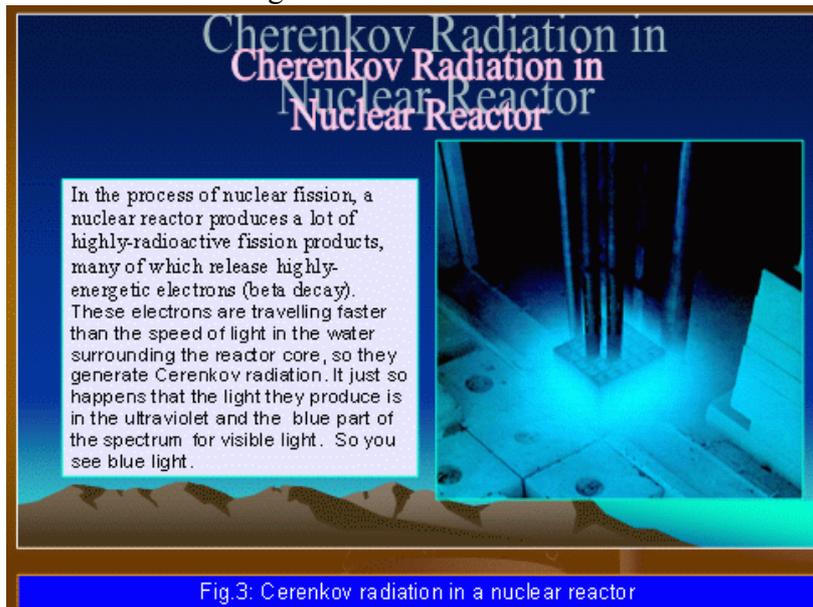

Fig.3: Cerenkov radiation in a nuclear reactor

Now, the Čerenkov radiation (ČR) is the subject of many studies related to the extension to the nuclear and hadronic media as well as to other coherent particle emission via Čerenkov-like mechanisms [14-45]. The generalized Čerenkov-like effects based on four fundamental interactions has been investigated and classified recently in [44]. In particular, this classification includes the nuclear (mesonic, $\gamma$, weak boson)- Čerenkov-like radiations as well as the high energy component of the coherent particle emission via (baryonic, leptonic, fermionic) Čerenkov-like effects. For some predictions on the *gravitational Čerenkov-like effects* see: M. Pardy, Phys. Lett. **B 336**, 362 (1994); Int. J. Theor. Phys. **34**, 951 (1995), **41**, 887 (2002); A. B. Balakin et al., Class. Quant. Grav. **18**, 2217 (2001).

## 2. Mesonic Čerenkov-like effects in hadronic media.

The idea that meson production in nuclear interactions may be described as a process similar to the Čerenkov radiation has considered by Wada (1949)[14], Ivanenko (1949)[15], Blohintev si Indenbom (1950)[16], Yekutieli (1959)[17], Czyz, Ericson, Glashow (1959-1960)[18-19], Smrz (1962)[20], D.B. Ion (1970)[21], D. B. Ion.(1971)[22], D.B. Ion and Nichitiu [23]. In 1971, in his Doctoral Thesis [21], D. B. Ion developed a general classical and quantum theory of the mesonic Čerenkov-like radiation in hadronic media.. Moreover, in the same thesis we introduced (in mondial premier) the barionic Čerenkov-like effects in nuclear and hadronic media. Then, we predicted completely all the properties of the (scalar and vector) mesonic Čerenkov-like radiation in the case when the mesonic refractive index is given by a single pole approximation. So, we obtained a good agreement with the integrated cross section of the single meson production in the hadronic collisions (see Refs. [21-23],[31-32],[38-39],[52]).

The investigations in this new field of scientific research as is gamma and meson spontaneous emissions via Čerenkov-like effects in nuclear media was continued by Prof. Dr. D.B. Ion and Prof. Dr. W.Stocker, at Munchen University, in a fruitfull collaboration on COSY-Jullich Projects. Then, we developed the classical and quantum theories for the *gamma* [29,33-34,36,40-42] and *mesonic* [30,37,40-41,43-44] *Čerenkov-like radiation in nuclear media*.):

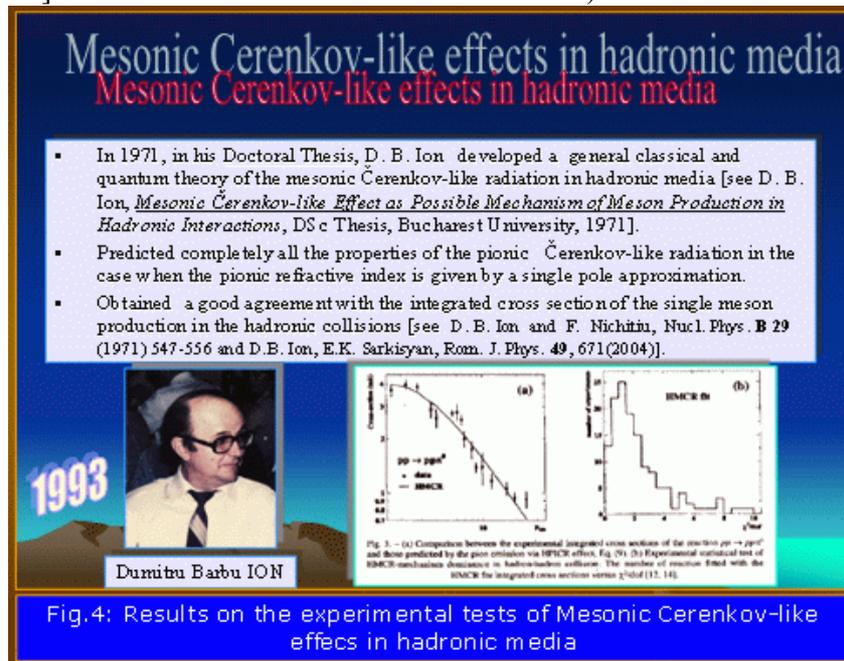

Now, it is important to note that in 1999, G.L.Gogiberidze, L.K. Gelovani and E.K. Sarkisyan [see Refs. [46-50]] performed the first experimental test of the pionic Čerenkov-like effect (NPICR) in Mg-Mg collisions at 4.3 GeV/c/nucleon. Then, they obtained a good agreement with the position and width of the first pionic Čerenkov-like band predicted by the profesors D.B.Ion and W. Stocker in ref. [12] (see Fig.5).

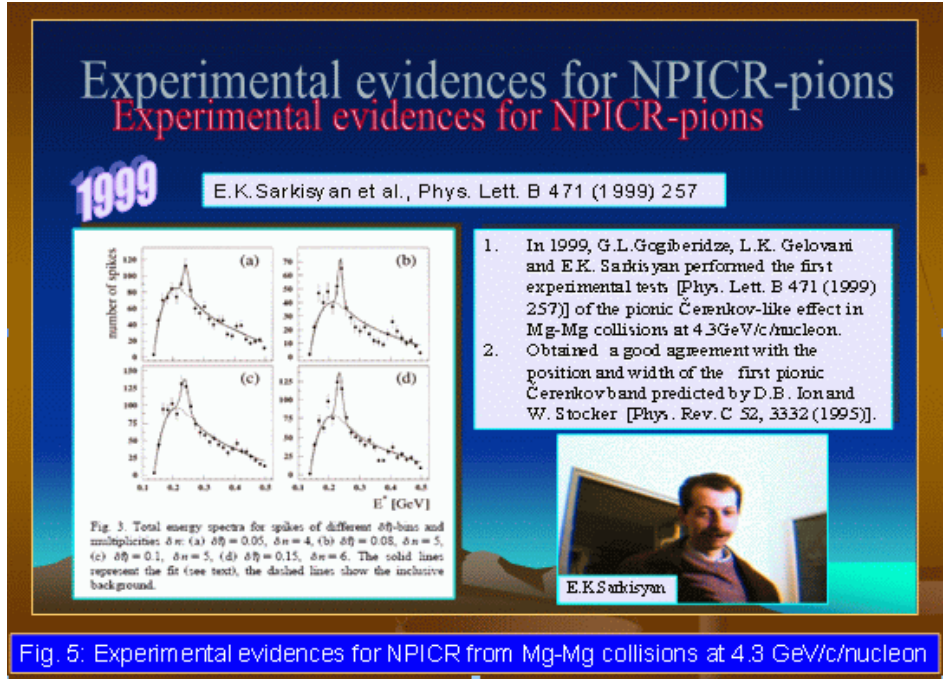

Fig. 5: Experimental evidences for NPICR from Mg-Mg collisions at 4.3 GeV/c/nucleon

### 3. Super-Čerenkov Radiation (SČR)

Recent theoretical investigations using the ČR correct kinematics lead us the discovery that ČR is in fact only the low energy component of a more general phenomenon called by us the ***Super-Čerenkov radiation*** (SČR) characterized by the Super-Čerenkov (SČR)-decay condition:

$$\cos\theta_{SC} = v_{xph}(E_x) \cdot v_{\gamma ph}(\omega_\gamma) \leq 1 \qquad (2)$$

where $v_{xph}(E_{xi})$ and $v_{xph}(E_{xf})$ are phase velocities of the charged particle in the initial and final states, respectively.

As we can see (see Fig. 9), the SČR-condition $\cos\theta_{x\gamma} = v_{\gamma ph}(\omega) \cdot v_{xph}(E_1) \leq 1$ is obtained in a natural way from the energy-momentun conservation law when the influence of medium on the propagation properties of the charged particle is also taken into account.

Indeed, the *super-Čerenkov relation* (2) can be easy proved by using the energy-momentum conservation law for the "decay" $1 \to 2 + \gamma$ (see Fig.8) to obtain the SČR-angles as they are given in Fig. 9. The signature of the SČR-effects are schematically described in Fig.10. Two components, corresponding to [low gamma energy and high gamma energy]-emmisions, with the orthogonal gamma polarizations, are clearly evidentiated

(a) Low $\gamma - energy$ sector or Čerenkov radiation sector (see Fig.10, upper part):

$$\cos\theta_{SC} = \cos\theta_{1\gamma} = v_{\gamma ph} \cdot v_{1ph} \Rightarrow \cos\theta_{1\gamma} = \frac{v_{\gamma ph}(\omega)}{v_1 \operatorname{Re} n_1} \leq 1 \qquad (3)$$

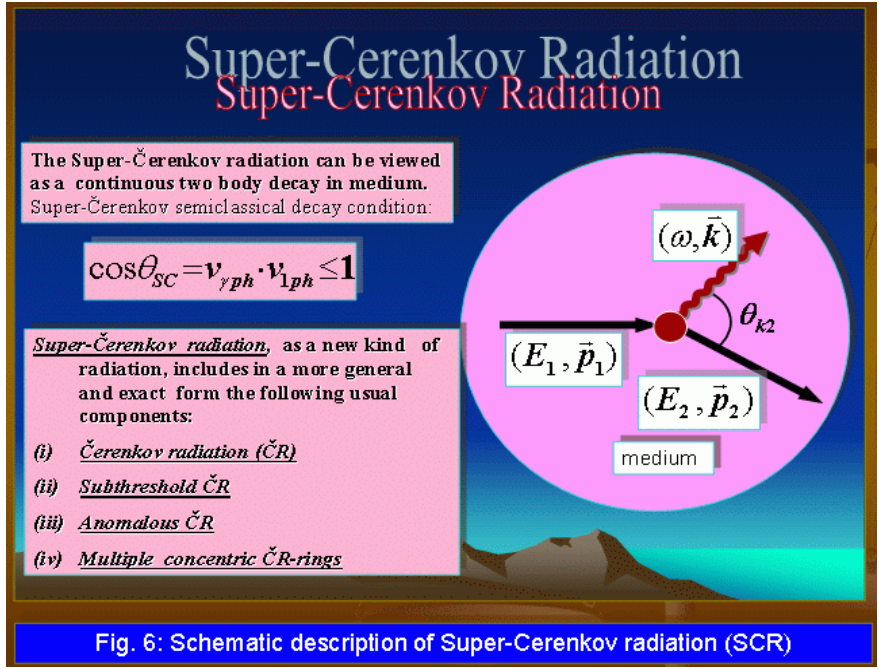

Fig. 6: Schematic description of Super-Cerenkov radiation (SCR)

(b) High $\gamma$ – *energy* or "*Source*" Čerenkov-like spontaneous bremsstrahlung sector (see Fig. 10), lower part:

$$\cos\theta_{SC} = \cos\theta_{12} = v_{1ph}.v_{2ph} \Rightarrow \cos\theta_{12} = \frac{v_{2ph}(E_2)}{v_1 \operatorname{Re} n_1} \leq 1 \qquad (4)$$

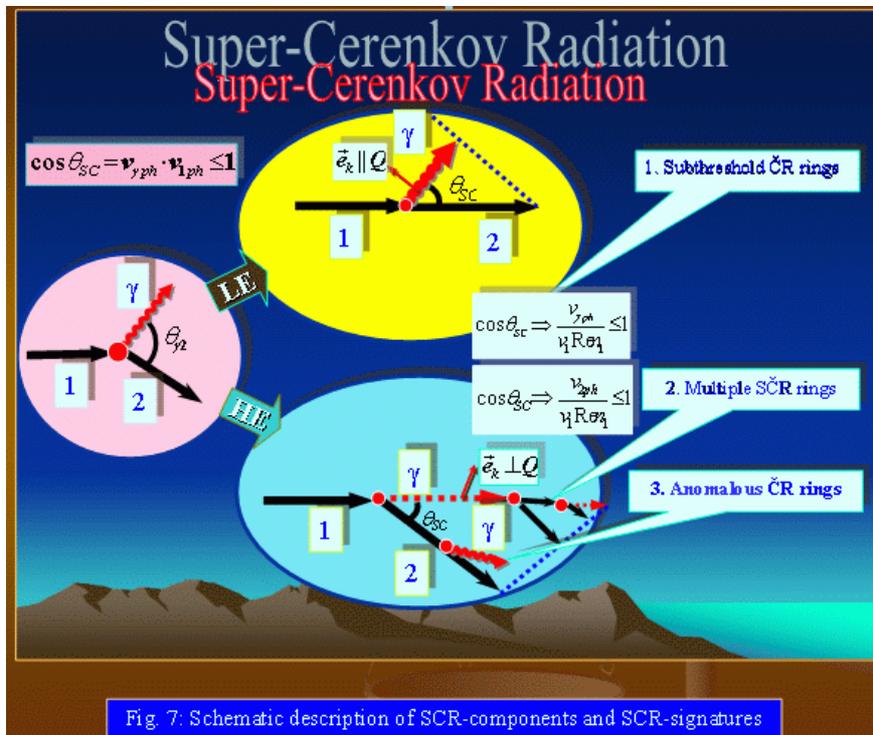

Fig. 7: Schematic description of SCR-components and SCR-signatures

Now, the subthreshold rings observed experimentally can also be interpreted as Super-Čerenkov signatures since in the ČR-sector $\theta_{SC} \equiv \theta_{2\gamma} = \theta_{1\gamma}$ (Fig.1a) the number of photons emitted in the (path and energy)-intervals (x, x+dx), $(\omega, \omega + d\omega)$ in an nonabsorbent medium will be given by [52-56]

$$\frac{d^2N}{dxd\omega}(SCR) \approx Z_{B_1}^2 \alpha \sin^2\theta_{1\gamma} = Z_1^2 \alpha[1 - v_{\gamma ph}^2 v_{1ph}^2] \tag{5}$$

where $\alpha = 1/137$ is the fine structure constant and $Z_1$ the electric charge of the particle 1. Indeed, it is easy to see that the Super-Čerenkov coherence condition (2-3) includes in a general and exact form the subthreshold Čerenkov-like radiation since: $v_1^{thr}(SCR) = v_1^{thr}(CR)/\mathrm{Re}\, n_1 \leq v_1^{thr}(CR)$ for $\mathrm{Re}\, n_1 \geq 1$.

## 4. Experimental Tests of Super-Čerenkov decay condition (2)

Čerenkov radiation is extensively used in experiments for counting and identifying relativistic particles in the fields of elementary particles, nuclear physics and astrophysics. A spherical mirror focuses all photons emitted at Čerenkov angle along the particle trajectory at the same radius on the focal plane as is shown in Fig.8. Photon sensitive detectors placed at the focal plane detect the resulting ring images in a Ring Imaging Čerenkov (RICH) detector. So, RICH-counters are used for identifying and traking charged particles. Čerenkov rings formed on a focal surface of the RICH provide information about the velocity and the direction of a charged particle passing the radiator. The particle's velocity is is related to the Čerenkov angle $\theta_C$ [or more exactly to the Super-Čerenkov angle $\theta_{SC}$] by the relations (1) [or (2)], respectively. Hence, these angles are determined by measuring the radii of the rings detected with the RICH. In Ref. [57] a $C_4F_{10}Ar(75:25)$ filled RICH-counter read out (by a 100-channel photomultipiler of $10 * 10$ cm² active area) was used for measurement in beams of the Čerenkov ring radii for electrons, muons, pions and kaons.

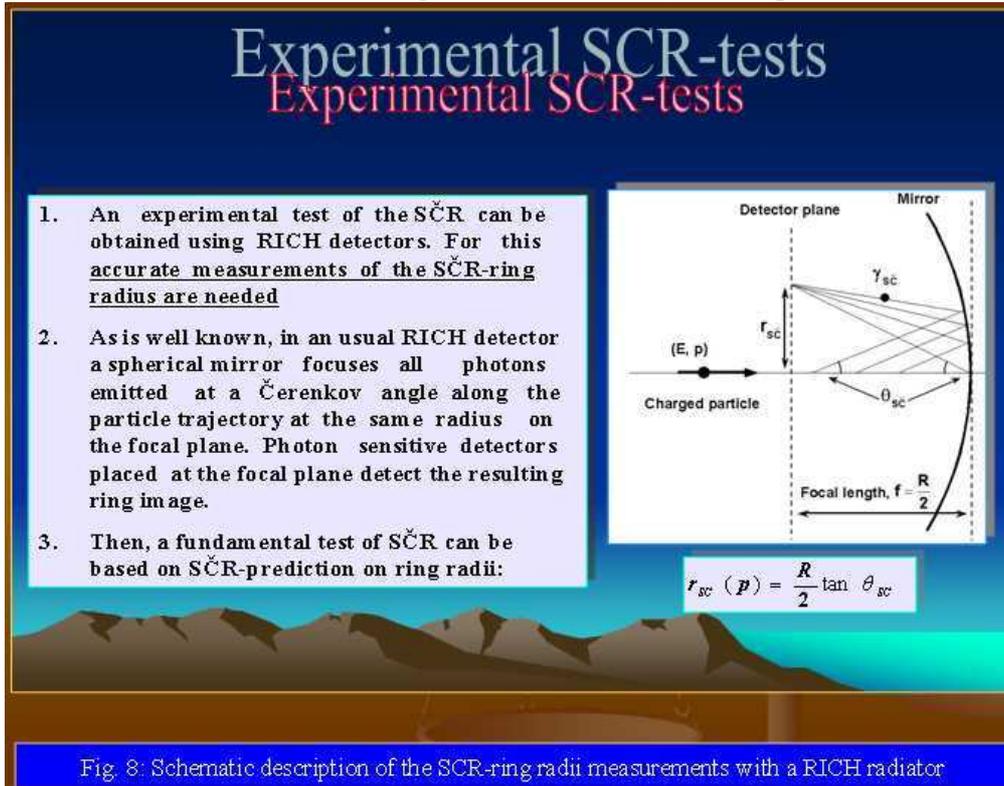

Fig. 8: Schematic description of the SCR-ring radii measurements with a RICH radiator

The results are compared in Fig. 9 with absolute prediction given by Čerenkov coherence condition (1) for the photon refractive index n=1.000113. In this case it is easy to see that the $\chi^2/dof \geq 10^5$.

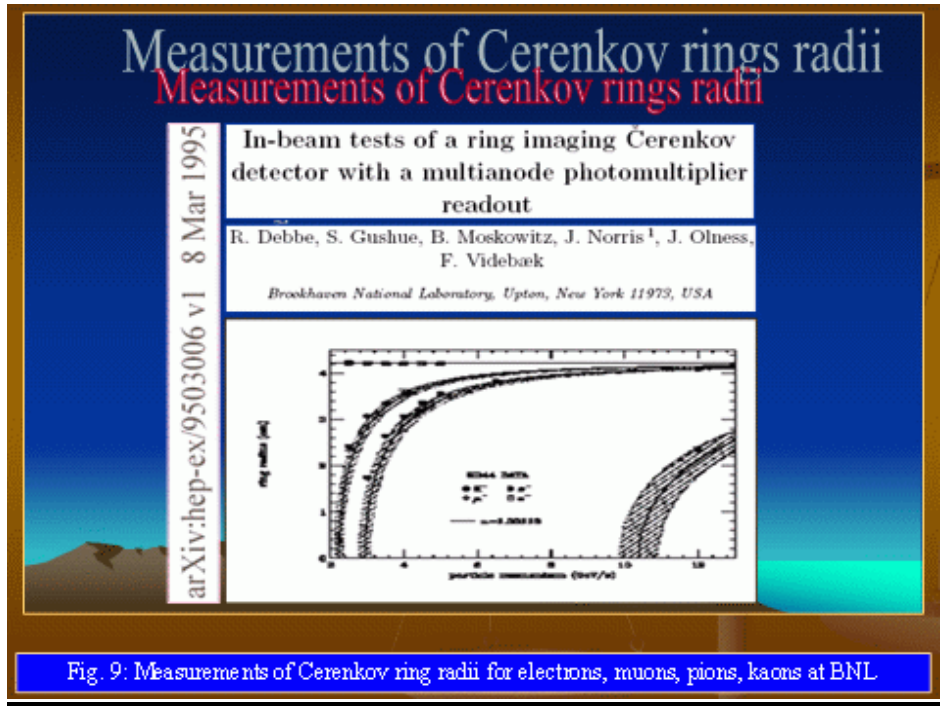

Fig. 9: Measurements of Čerenkov ring radii for electrons, muons, pions, kaons at BNL

A detailed tests of the supercoherence condition (2) is presented in Figs.10-12 (see [54-56]).

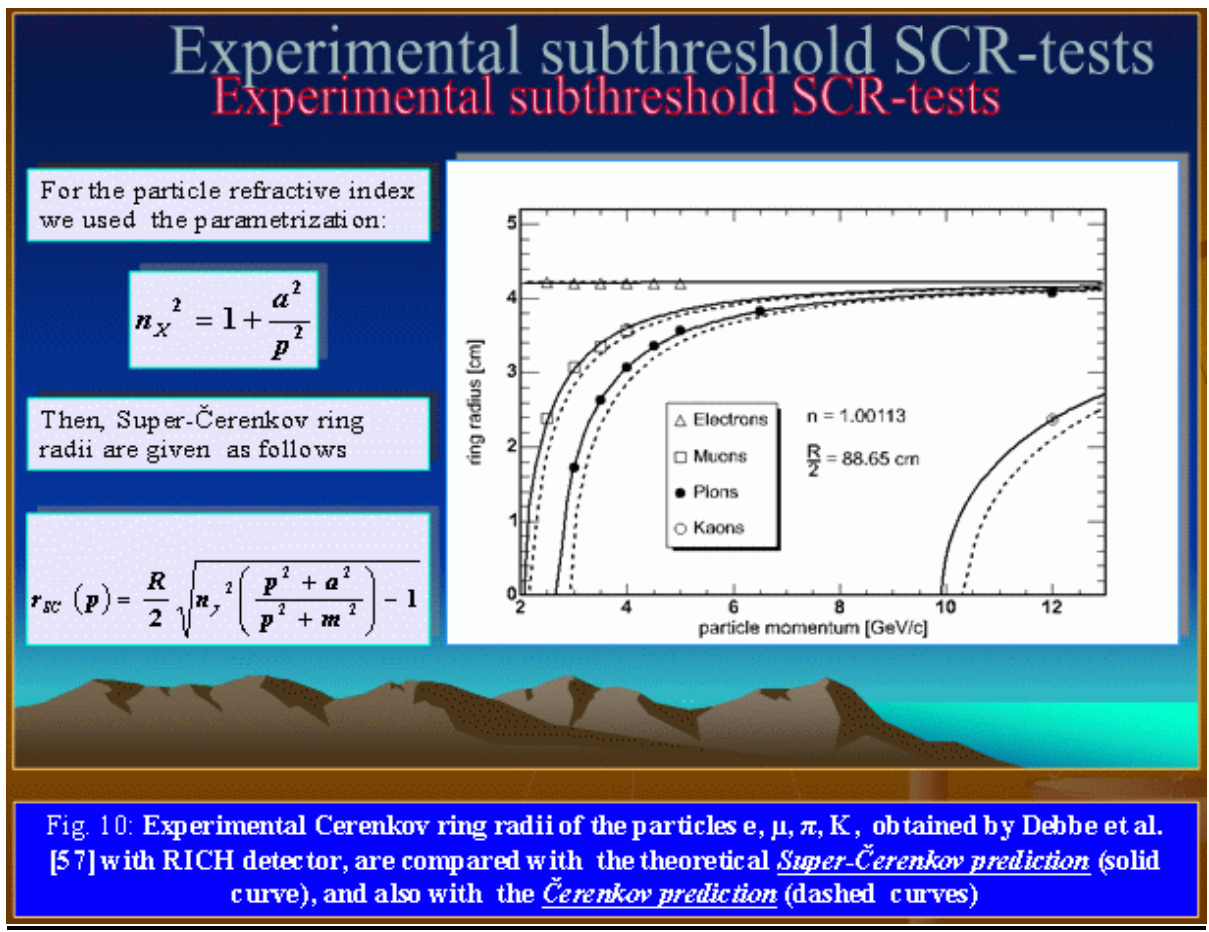

Fig. 10: Experimental Čerenkov ring radii of the particles e, μ, π, K, obtained by Debbe et al. [57] with RICH detector, are compared with the theoretical *Super-Čerenkov prediction* (solid curve), and also with the *Čerenkov prediction* (dashed curves)

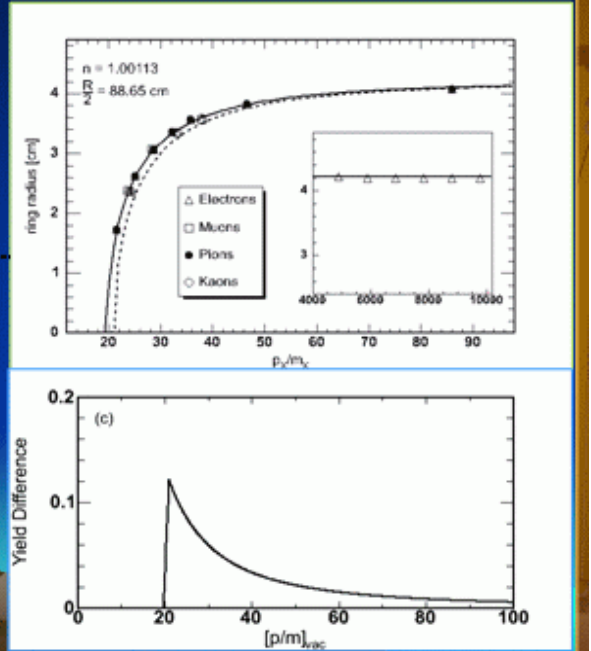

Fig. 11: Scaling of SCR-rings radii

Fig. 12: The results of SCR-fits

## 4. Anomalous Čerenkov Rings as important signature of Super-Čerenkov Radiation

The Čerenkov radiation caused by relativistic lead ions was studied at SPS CERN [58]. The Čerenkov gas detector was placed in the H2 beam line in the location of NA49 set-up between the large time-projection chamber and the barrel hadron calorimeter. The general layout of the detector is shown in Fig. 13.

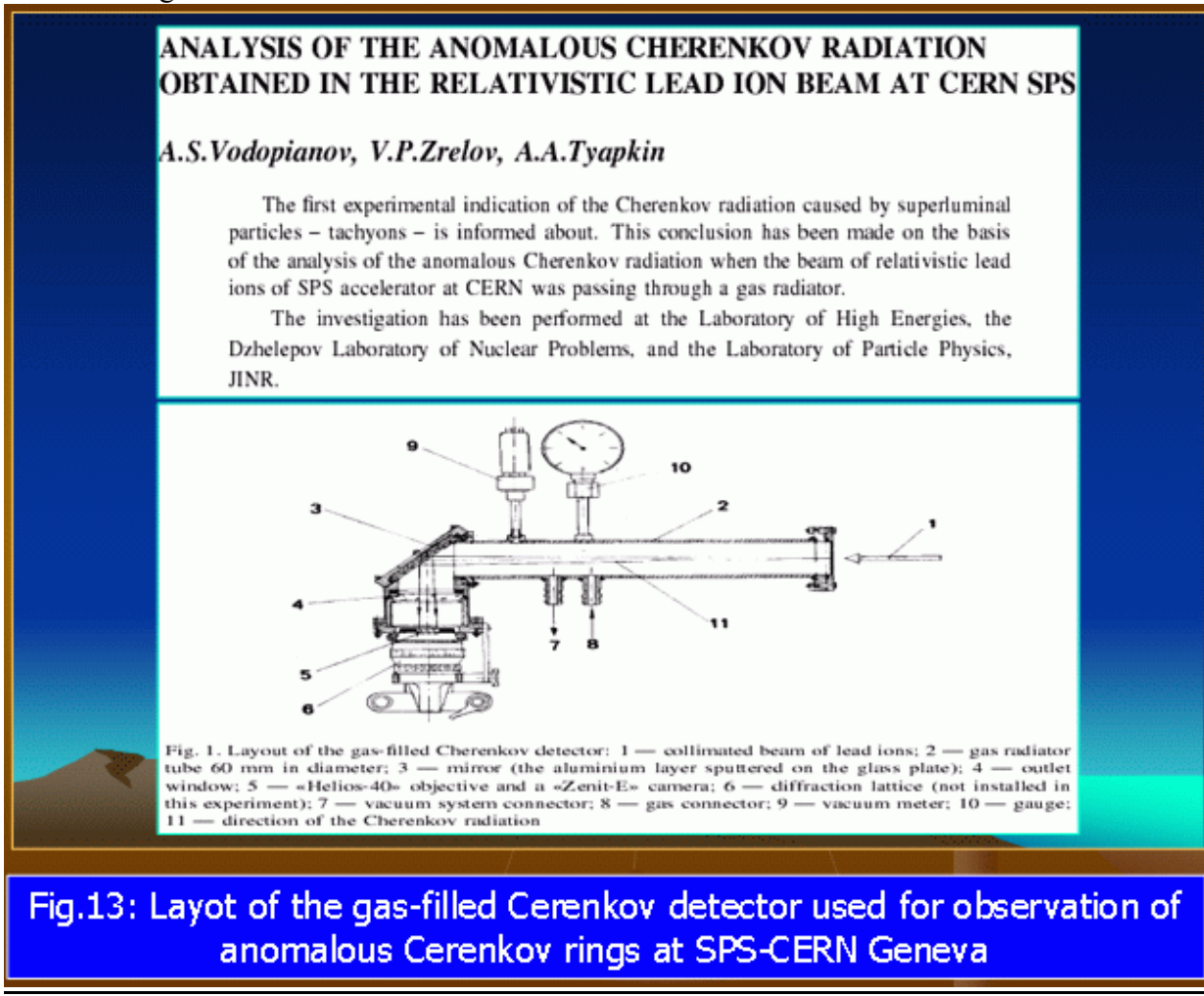

Fig.13: Layot of the gas-filled Cerenkov detector used for observation of anomalous Cerenkov rings at SPS-CERN Geneva

A beam of $^{208}Pb^{82+}$ ions with the energy of 157.7 A GeV was going along the axis of the Čerenkov detector. The Čerenkov light emitted in the radiator (its length along the optical axis is 405 mm) got into the objective of a photocamera after its refection in the mirror inclined under 45 degree angle relative to the axis of the radiator. The inner surface of the detector tube was covered by soot to avoid light refections (see Fig. 13 and original paper [58] for other details).

A bright narrow ring of the Čerenkov radiation seen on the picture (see Fig. 14) is caused by relativistic lead ions. Besides, in this picture we have found hardly noticeable narrow Čerenkov radiation rings of the particles fying out under small angles to the direction of the beam. The calculation of the velocity of these particles has shown that it corresponds to those of particles moving faster than the light velocity in the vacuum.

The conclusion of Vodopianov et al [58] is that the large Čerenkov radiation ring, shown with an arrow in Fig. 14, corresponds to a tachion velocity approximately equal to $\beta \approx 1.0008$. The ring diameter of its radiation is approximately two times larger than the ring diameter of the proton radiation at the velocity of motion $\beta \to 1$.

Totally, seven rings of the anomalous Čerenkov radiation have been found in the three photos [58]. The locations of rings in these photographs are shown in Fig.15. The general numeration for them is given as well. Thick circles in photoshots *a,b, c* give the location of the Čerenkov radiation ring of the lead ion beam in the air.

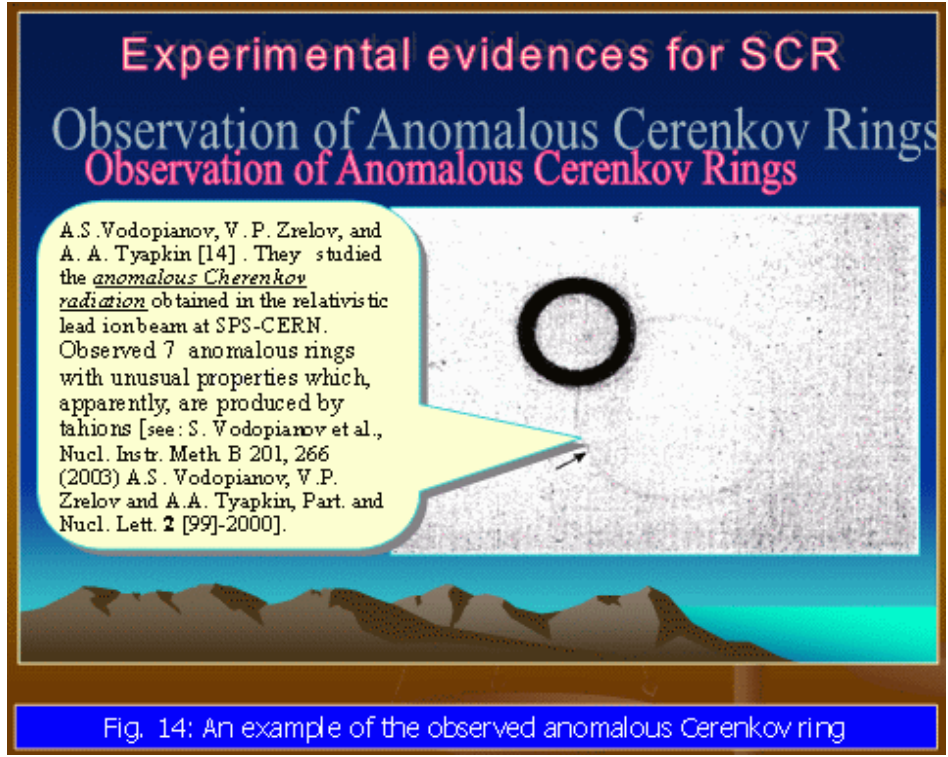

Fig. 14: An example of the observed anomalous Cerenkov ring

The rings were analysed by Vodopianov et al. [58] using the standard approximation expression :

$$\cos\theta = 1/n_\gamma v \qquad (6)$$

insted with the true complete formula

$$\cos\theta = \frac{1}{n_{Pb^+}(E)n_\gamma(\omega)v} \qquad (7)$$

where $n_{Pb^+}(E)$ is the refractive index of the lead ions $^{207}_{82}Pb^+$ in medium. Therefore, the authors of Ref. [58] obtained in their table the values for $\beta n_{Pb^+}$ instead the values of $\beta$.

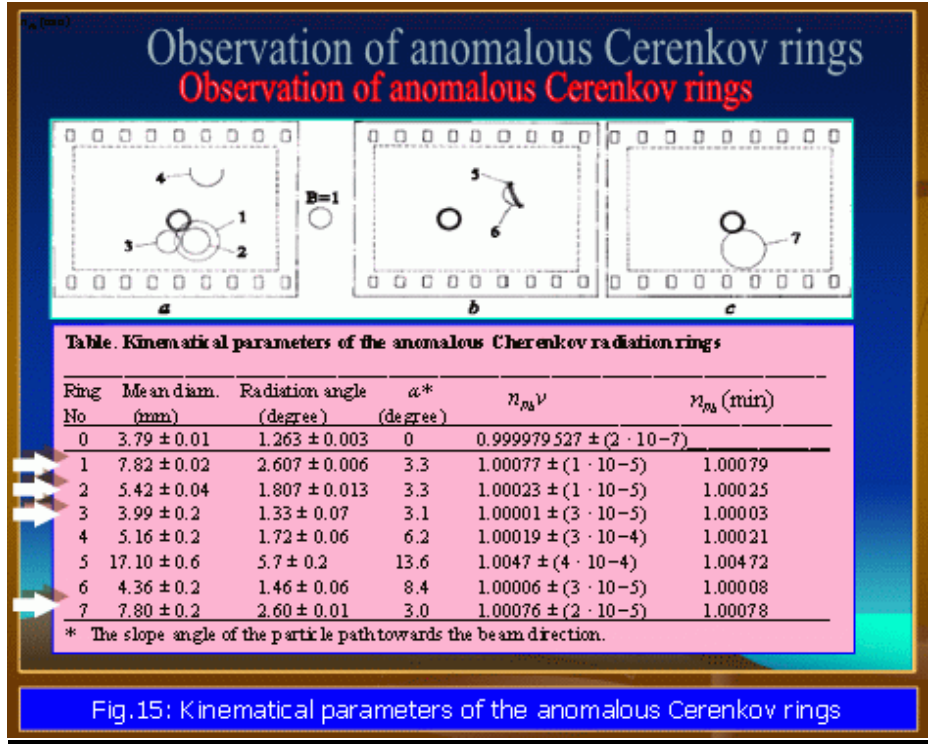

Fig.15: Kinematical parameters of the anomalous Cerenkov rings

Therefore, the observation of the anomalous Čerenkov ring can be interpreted as one of the most important signature of the HE-component of Super-Čerenkov radiation (see Fig. 7) produced by lead ions in the radiator medium.

## 5. Conclusions

The main results and conclusions can be summarized as follows:

1. The *Super-Čerenkov* (SČR) phenomenon, can be viewed as a two-body decays in medium and is expected to take place when the final particles satisfy the dual super-coherence condition:

$$\cos\theta_{SC} = v_{\gamma ph} \cdot v_{xph} \leq 1$$

2. The Super-Čerenkov Radiation phenomenon includes in an unified way :
☯ (i) Gamma-Čerenkov radiation including subthreshold ČR (see Fig.7, *LE-component*)
☯ (ii) "Particle source" Čerenkov-like effect (see Fig.10, *HE-component*)
☯ (iii) Anomalous Čerenkov radiation (secondary SČR-rings) (see Figs.7,14)
☯ (iii) Secondary multiple SČR-rings phenomena (see Fig.7)

3. The experimental test of of this SČR-coherence relation, near the usual Čerenkov threshold, is performed by using the data of Debbe et al. [57] on Čerenkov ring radii of electrons, muons, pions and kaons in a RICH detector. The results on this experimental test of the super-coherence conditions are presented in Fig 10-12. These SČR-predictions are verified experimentally with high accuracy: χ2/dof =1.47. The relative intensity of the SČR-phenomenon at Čerenkov threshold, inferred from data of Debbe et al. [57], is estimated to be about 13% (see Figs. 11-12) for all particles: e, μ, π, K, etc;

4. The observation of the anomalous Čerenkov rings [58] is interpreted as one of the most important signature of the HE-component of Super-Čerenkov radiation (see Fig.7) produced by lead ions in the radiator medium.
5. The RICH detector can be used to investigate the influence of medium on the particle propagation properties in medium. The refractive properties of electrons, muons, pions, in a radiator $C_4F_{10}Ar$ are obtained. So, it is shown that the refractive indices of the particles in medium are also very important for the RICH detectors, especially at low and intermediate energies.

Finally, we remark that new and accurate experimental measurements of the Čerenkov ring radii, as well as for the anomalous HE-component of SČR are needed.